\documentstyle[aps,prd,preprint,epsfig]{revtex}

\newcommand{\beq}{\begin{equation}}
\newcommand{\eeq}{\end{equation}}

\begin{document}
\preprint{LPT-ORSAY 03-32}
\draft
\title{Gravity of superheavy higher-dimensional global defects}

\author{Inyong Cho\footnote{Electronic address: 
Inyong.Cho@th.u-psud.fr}}
\address{Laboratoire de Physique Th\'eorique,
Universit\'e Paris-Sud, B\^atiment 210, F-91405   
Orsay CEDEX, France}
\author{Alexander Vilenkin\footnote{Electronic address: 
vilenkin@cosmos.phy.tufts.edu}}
\address{Institute of Cosmology, Department of Physics and Astronomy,
Tufts University, Medford, Massachusetts 02155, USA}

\date{\today}

\maketitle

\begin{abstract}
Numerical solutions of Einstein's and scalar-field equations are found
for a global defect in a higher-dimensional spacetime.  The defect has
a $(3+1)$-dimensional core and a ``hedgehog'' scalar-field
configuration in $n=3$ extra dimensions.  For sufficiently low
symmetry-breaking scales $\eta$, the solutions are characterized by a
flat worldsheet geometry and a constant solid deficit angle in the
extra dimensions, in agreement with previous work.  For $\eta$ above
the higher-dimensional Planck scale, we find that static-defect
solutions are singular.  The singularity can be removed if the
requirement of staticity is relaxed and defect cores are allowed to
inflate.  We obtain an analytic solution for the metric of such
inflating defects at large distances from the core.  The three extra
dimensions of the nonsingular solutions have a ``cigar'' geometry.
Although our numerical solutions were obtained for defects of
codimension $n=3$, we argue that the conclusions are likely to apply
to all $n\geq 3$.
\end{abstract}

\vspace{0.5in}

\section{Introduction}
According to the ``brane-world'' picture, our universe is a
$(3+1)$-dimensional brane floating in a higher-dimensional bulk
spacetime~\cite{Rubakov,Arkani,RS}. The branes can be thought of as
fundamental $D$-branes, or they can arise as topological defects in a
higher-dimensional field theory.  It may well be that $D$-branes could
also eventually find some dual field-theory description.  In the
simplest, codimension-one models, the brane can be pictured as a
domain wall propagating in a 5D spacetime~\cite{Rubakov,Wall}.
Higher codimensions with both gauge and global defects have also been
considered.  In particular, models have been discussed where the field
configuration in the directions orthogonal to the brane is that of a
cosmic string~\cite{String}, a monopole~\cite{Olasagasti,Monopole}, or
a texture~\cite{Texture}.  In these models, the bulk curvature produced
by the defects plays a major role in localizing gravity on the brane
and in solving the mass-hierarchy problem.

The solutions of Einstein's equation describing higher dimensional
defects with a global charge found in Ref.~\cite{Olasagasti} are very
simple.  In the absence of a bulk cosmological constant, and for
codimension $n>2$, the spacetime is flat along the brane worldsheet
and is characterized by a constant solid angle deficit in the
transverse directions.  The angle deficit grows as the symmetry
breaking scale $\eta$ of the defect is increased, and eventually
consumes the entire solid angle at some critical value $\eta_c$, which
is comparable to the higher-dimensional Planck scale.  For
$\eta=\eta_c$, the transverse metric becomes that of a cylinder.

The solution of Ref.~\cite{Olasagasti} were obtained in the asymptotic
region outside the defect core.  In the core region, analytic
solutions cannot be found, and one of the purposes of the present
paper is to check numerically that the asymptotic solutions 
of Ref.~\cite{Olasagasti} can be matched to appropriate core solutions.

Another important question is what happens if $\eta$ is greater than
$\eta_c$.  One of the motivations for looking into this regime is the
interesting work by Dvali, Gabadadze, and Shifman~\cite{DGS},
suggesting that brane-world models of codimension $n>2$,
with a very low value of the bulk Planck scale, can provide a solution
to the cosmological constant problem.  It was conjectured in Ref.~\cite{DGS}
that defect solutions in such models may exhibit de Sitter inflation
of the worldsheet, with the expansion rate {\it inversely}
proportional to the brane tension.  The very low-accelerated expansion
rate observed today could then be obtained with a very large-brane
tension.  

The main purpose of the present paper is to solve Einstein's and
scalar-field equations for super-critical global defects.
Specifically, we consider a codimension-3 global monopole.  In the
following Section, we introduce the model and review the asymptotic
solutions of Ref.~\cite{Olasagasti}.  The numerical solutions assuming
a static (noninflating) worldsheet are presented in Sec.~III, for
both sub- and super-critical regimes.  We find that super-critical
solutions have a curvature singularity at a finite distance from the
monopole core.  Inflating defect solutions are discussed in Sec.~IV, 
where we show that the expansion rate $H$ can be adjusted so that
the singularity is removed.  We interpret the resulting nonsingular
geometries as the physical defect solutions.  Our conclusions are
summarized in Sec.~V.

\section{Field equations and asymptotic solutions}
The action for our model is
\begin{equation}
{\cal S} = \int d^7x \sqrt{-g}\left[
{{\cal R} \over  2\kappa^2} 
-{1\over 2}\partial_A\phi^a\partial^A\phi^a
-{\lambda\over 4}(\phi^a\phi^a-\eta^2)^2
\right]\,,
\label{eq=action}
\end{equation}
where $\kappa^2=1/M^{2+n}$, $M$ is the 7D Planck mass, and $\eta$ is
the symmetry-breaking scale.  The scalar-triplet field $\phi^a$ is
assumed to be in a global monopole ``hedgehog'' configuration in $n=3$
extra transverse dimensions, $\phi^a = \phi(r)\hat{x}^a$.

The general static form of the seven dimensional metric with spherical
symmetry in the extra three dimensions is
\begin{equation}
d s^2 = B^2(r)d\bar{s}_4^2+ dr^2 + C^2(r)r^2d\Omega_2^2\,,
\label{eq=metric}
\end{equation}
where $d\bar{s}_4^2$ represents the  4D world-volume metric. 
The corresponding Einstein equations are
\begin{eqnarray}
-G^\mu_\mu &=& -3{B'' \over B} -3\left({B'\over B}\right)^2
-6{B' \over Br} -6{B'C'\over BC} -2{C''\over C}
-\left({C'\over C}\right)^2 -6{C'\over Cr}
+{1\over C^2r^2}-{1\over r^2}+{1\over 4}{\bar{R}^{(4)}\over B^2} 
\nonumber \\
{} &=& \kappa^2\left[ {\phi'^2 \over 2} +{\phi^2 \over C^2r^2} 
+{\lambda\over 4}(\phi^2-\eta^2)^2
\right] \,,
\label{eq=Gtt} \\
-G^r_r &=& -6\left({B'\over B}\right)^2
-8{B' \over Br} -8{B'C'\over BC}
-\left({C'\over C}\right)^2 -2{C'\over Cr}
+{1\over C^2r^2}-{1\over r^2}+{1\over 2}{\bar{R}^{(4)}\over B^2} 
\nonumber \\
{} &=& \kappa^2\left[ -{\phi'^2 \over 2} +{\phi^2 \over C^2r^2} 
+{\lambda\over 4}(\phi^2-\eta^2)^2
\right] \,,
\label{eq=Grr}\\
-G^{\theta_i}_{\theta_i} &=& 
-4{B'' \over B} -6\left({B'\over B}\right)^2
-4{B' \over Br} -4{B'C'\over BC} -{C''\over C}
-2{C'\over Cr}
+{1\over 2}{\bar{R}^{(4)}\over B^2} 
\nonumber \\
{} &=& \kappa^2\left[ {\phi'^2 \over 2}
+{\lambda\over 4}(\phi^2-\eta^2)^2
\right]\,, 
\label{eq=Gthth}
\end{eqnarray}
where $\bar{R}^{(4)}$ is the Ricci scalar of the 4D world volume.

The field equation for the scalar field is
\begin{equation}
\phi'' + 2\left( 2{B'\over B} +{C'\over C}+{1\over r}\right)\phi'
-2{\phi \over C^2r^2} -\lambda\phi (\phi^2-\eta^2) =0\,.
\label{eq=phi}
\end{equation}

The solution obtained in Ref.~\cite{Olasagasti}, assuming a flat 4D
world-volume, $\bar{R}^{(4)}=0$, and $\phi =\eta$ in the asymptotic
region, is of the form
\begin{equation}
ds^2 = \eta_{\mu\nu}dx^\mu dx^\nu +dr^2
+(1-\kappa^2\eta^2)r^2d\Omega_2^2\,.
\label{itsaso}
\end{equation}
The solid angle deficit is $\Delta\Omega=4\pi\kappa^2\eta^2$, and 
the critical symmetry-breaking scale is 
\beq
\eta_c=1/\kappa.
\eeq
It was shown in Ref.~\cite{Olasagasti} that for this value of $\eta$,
the Einstein equations have a cylindrical solution
\beq
ds^2 = \eta_{\mu\nu}dx^\mu dx^\nu +dr^2
+C_0^2d\Omega_2^2\,,
\label{cigar}
\end{equation}
where $C_0$ is an arbitrary constant.  The expectation was that this
constant radius of the cylinder can be determined by matching to an
appropriate-core solution, so that the full 3D geometry will be that
of a ``cigar''.  In the next section, we shall see that {\it (i)} this
expectation for the critical solution is not supported by the
numerical analysis, but {\it (ii)} cigar-type geometries do
nonetheless make their appearance in super-critical monopole
solutions.

\section{Static solutions}

We numerically solved the Einstein and scalar-field equations
(\ref{eq=Gtt})-(\ref{eq=phi}) with boundary conditions $B(0)=C(0)=1$,
$B'(0)=C'(0)=0$, $\phi (0)=0$, and $\phi=\eta$ at large $r$.  For
static solutions, we assumed a flat worldsheet, $d{\bar s}_4^2 =
\eta_{\mu\nu}dx^\mu dx^\nu$, and ${\bar R}^{(4)}=0$.

\begin{figure}
\begin{center}
\epsfig{file=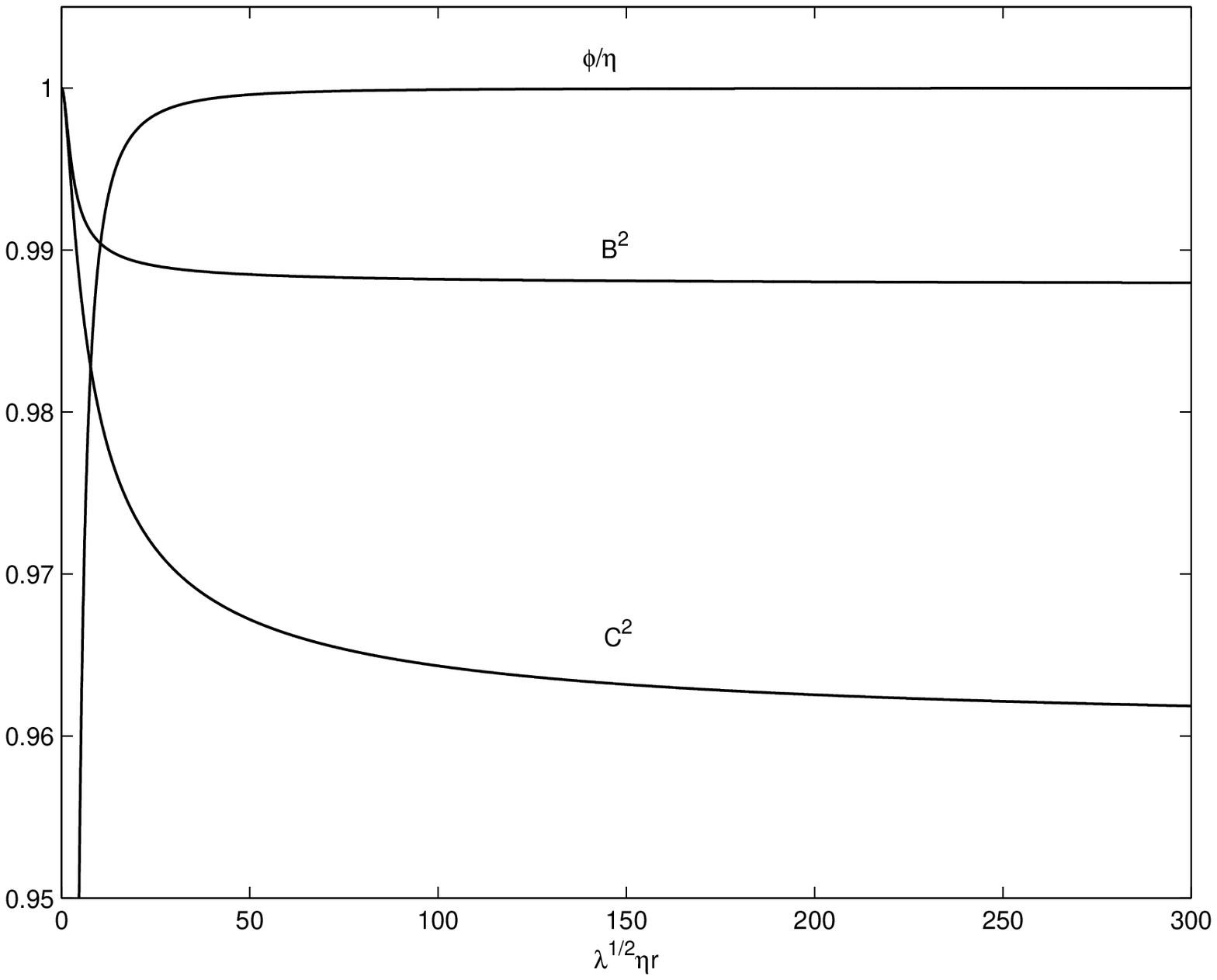,width=5in}
\end{center}
\vspace{0.5in}
\caption{
Scalar and gravitational fields of a sub-critical monopole,
$\kappa\eta = 0.2$. 
The scalar field rapidly approaches $\eta$.
(The lower part of the graph is not shown.)
The metric coefficients $B^2$ and $C^2$ 
approach constants. $C^2$ exhibits the expected
solid angle deficit in the asymptotic region.
}
\label{fig=regBC}
\end{figure}

The solutions for the scalar field $\phi$ and the metric of a
sub-critical monopole ($\kappa\eta = 0.2$) are shown in
Fig.~\ref{fig=regBC}.  As expected, $\phi$ rapidly approaches $\eta$,
and the metric coefficients $B$ and $C$ approach constants at large
$r$.  The asymptotic value of the angle-deficit coefficient $C$ is in
agreement with Eq.~(\ref{itsaso}).  As $\eta$ gets closer to
$\kappa^{-1}$, the approach to this asymptotic value becomes
increasingly slow.  

In the critical case, $\kappa\eta=1$, $C \to 0$ as $r\to\infty$.
However, the radius of the extra dimensions $Cr$ is a growing function
of $r$ (see Fig.~2).  Thus, the asymptotic behavior of the
critical-monopole solution is not described by the cylindrical
metric~(\ref{cigar}).  With $\phi \approx \eta$, we were able to find
a power-law analytic solution of Einstein equations,
\begin{eqnarray}
B &\propto & r^{(2-\sqrt{10})/12} \approx r^{-0.097}\,,\\
C &\propto & r^{(-5+\sqrt{10})/6} \approx r^{-0.306}\,,
\end{eqnarray}
which gives a good fit to our numerical solution at large $r$.

In the super-critical case, $\kappa\eta>1$, we find that static
solutions develop a curvature singularity at a finite distance from
the defect core.  When $\eta$ is only slightly above the critical
value, this distance can be very large.  The solution initially
behaves as the ``cigar'' solution~(\ref{cigar}), but eventually the
radius of extra dimensions blows up, $C\to\infty$, and the metric
coefficient $B$ plummets to zero at the same finite value of $r=r_s$.
This behavior is illustrated in Fig.~\ref{fig=singBC} for $\kappa\eta=1.06723$.
The nature of this singularity is similar to that of a static 
4D global string~\cite{Sstring}.

As the symmetry-breaking scale $\eta$ is increased, the singularity
gets closer to the core of the monopole.  We performed numerical
calculations for a number of values of $\eta$; the resulting
dependence $r_s(\eta)$ is plotted in Fig.~\ref{fig=rs}.  For $\eta
\gtrsim 1.2$, the singularity gets too close to the center, so that
the scalar field does not fully relax to its asymptotic value, $\phi =
\eta$, before the singularity forms.

\section{Inflating solutions}
It is well known that the singularity of some static-defect solutions
can be removed by relaxing the requirement of staticity and allowing
the defect worldsheet to inflate.  Examples are domain
walls~\cite{AV83,Ipser} and global strings~\cite{Gregory,Santos} in
$(3+1)$ dimensions.  It is, therefore, reasonable to check if the same
strategy works in higher dimensions.

The 4D metric of an inflating defect worldsheet is the de Sitter metric,
\begin{equation}
d\bar{s}_4^2 = -dt^2 +e^{2Ht}d\text{x}^2\,,
\label{deSitter}
\end{equation}
and the 4D curvature scalar is $\bar{R}^{(4)}=12H^2$.  We solved the
equations numerically for different values of $H$, and indeed, for
values of $\eta$ above, but not too far above the critical, we were
able to adjust $H$ so that the singularity is removed.  Two
super-critical, nonsingular solutions are shown in Fig.~\ref{fig=flat}.  

We see that the extra dimensions in these solutions have a ``cigar''
geometry, with $Cr\to const$ at large $r$, and that the scalar
field quickly approaches a constant value, which, however, is somewhat
below $\eta$.  The latter fact can be easily understood from the field
equation~(\ref{eq=phi}) for $\phi$.  With $\phi'=0$ and
$\sqrt{\lambda}\eta Cr\equiv C_0={\rm const}$, it gives
\beq
{\phi^2\over{\eta^2}}=1-{2\over{C_0^2}}.
\eeq
We were able to find an exact analytic solution of our system of equations,
which describes the asymptotic behavior of these
nonsingular solutions:
\begin{eqnarray}
\phi &=& \phi_0 = {\sqrt{5-\kappa^2\eta^2} \over 2\kappa}\,,
\label{eq=phi0}\\
\sqrt{\lambda}\eta B &=& {H\over k}\sin (\sqrt{\lambda}\eta kr)\,,
\label{eq=B}\\
\sqrt{\lambda}\eta Cr &=& C_0 = \sqrt{8\over 5} {\kappa\eta \over 
\sqrt{\kappa^2\eta^2 -1}}\,,
\label{eq=C0}
\end{eqnarray}
where $k = \sqrt{5/128}(\kappa^2\eta^2 -1)/\kappa\eta$.

In order to solve the equations numerically, we had to impose the
boundary condition $\phi=\phi_0$ at large $r$.  The actual sequence of
steps that led to the solution was that, after experimenting with
numerical solutions for different values of $H$ with our old boundary
condition, $\phi(r\to\infty)=\eta$, we guessed that the nonsingular
solution should have a cigar geometry.  We then found the analytic
solutions~(\ref{eq=phi0})-(\ref{eq=C0}) and used~(\ref{eq=phi0}) to set
the boundary condition.

In the asymptotic solution~(\ref{eq=B}), the expansion rate $H$ can be
absorbed by rescaling the time coordinate $t$ in Eq.~(\ref{deSitter});
then the solution takes the form
\begin{equation}
(\lambda\eta^2) ds^2 = k^{-2}(\sin^2\chi d\bar{s}_+^2 + d\chi^2)
+ C_0^2d\Omega_2^2\,,
\label{itsaso2}
\end{equation}
where $\chi=\sqrt{\lambda}\eta kr$ and $d\bar{s}_+^2$ is the 4D de
Sitter metric with $H=1$.  The above metric is essentially the same as
the one found in Ref.~\cite{Olasagasti} as a solution for $\phi =\eta$
and a positive bulk cosmological constant.  In our model, the role of
the cosmological constant is played by the scalar-field potential,
$V(\phi_0)>0$.

The solution~(\ref{itsaso2}) has an apparent singularity at
$\chi=\pi$, but as it was noted in Ref.~\cite{Olasagasti}, the first two
terms in Eq.~(\ref{itsaso2}) describe a 5D de Sitter space.  The 5D
inflation rate is $H_5=\sqrt{\lambda}\eta k$, and the surface
$\chi=\pi$ corresponds to the de Sitter horizon.  This shows that the
space outside the monopole core inflates not only along the 4D
worldsheet, but also in one of the transverse directions, while two of
the extra dimensions remain compactified in a sphere.  This behavior
is reminiscent of the topological-inflation scenario~\cite{V94,Linde94}, 
where inflation occurs in the cores of topological
defects in 4D.  There are, however, important differences.  First,
the defect core has a fixed radius and is not expanding in the radial
direction, and second, two of the extra dimensions remain
compactified, also at a fixed radius.

Although the 4D expansion rate $H$ drops out of the asymptotic
solution, it is a meaningful parameter for the full spacetime.  With
our boundary conditions at $r=0$, it gives the inflation rate in the
core of the monopole.  As we saw in the preceding section, the static
$H=0$ solutions with $\kappa\eta > 1$ are singular.  As $H$ is
increased, the singularity gets milder, and at a certain value of $H$
it disappears completely, resulting in a cigar geometry.  We interpret
the corresponding solution as the physical solution describing the
spacetime of a superheavy defect.  

Our conjecture is that for each $\eta$, only one particular value of
$H$ gives a nonsingular solution.  Of course, it is impossible to find
this precise value numerically, and therefore, all super-critical
numerical solutions that we considered eventually developed a
singularity.  The nature of the singularity is similar to that of a
static solution in Fig.~\ref{fig=singBC}, with $B\to 0$ and $C\to\infty$.  
By adjusting the value of $H$, we were able to shift the onset of the
singularity to larger and larger radii, so that the solution got
closer and closer to the asymptotic solutions
(\ref{eq=phi0})-(\ref{eq=C0}).  In the end, however, $C(r)$ invariably
started to diverge, at some point very close to the would-be de Sitter
horizon, $r\approx \pi/\sqrt{\lambda}\eta k$.  We believe that our
results are very suggestive, but they do not, of course, constitute a
proof that all values of $H$ except one yield singular solutions.  
It may be possible to provide a proof using the dynamical-systems
method employed in Ref.~\cite{Gregory}.

Our numerical results indicate that, for the nonsingular solution, $H$
grows as we increase $\eta$, and the growth is nearly linear, at least,
for $\eta$ sufficiently close to $\eta_c$ (see Fig.~\ref{fig=H}).  
Above some
value, $\kappa\eta > 1.1$, it is difficult to get a cigar solution
numerically.  
As $\eta$ ($H$) increases, 
$Cr$ becomes more wiggly in the asymptotic region 
unless $\eta$ and $H$ are tuned very delicately
to get a cigar solution.
This causes numerical
difficulties in getting a cigar solution for large $\eta$.

From Eq.~(\ref{eq=phi0}) we see that the asymptotic solution exists
only for $1 < \kappa\eta <\sqrt{5}$.  At the end of this interval,
$\kappa\eta =\sqrt{5}$, the scalar field is in the unbroken-symmetry
state, $\phi =0$.  Then the source term in the Einstein's equation is pure
vacuum, $T_\mu^\nu=V(0)\delta_\mu^\nu$, and the solution
(\ref{itsaso2}) takes the form  
\begin{equation}
(\lambda\eta^2) ds^2 = 8(\sin^2\chi d\bar{s}_+^2 + d\chi^2)
+ 2d\Omega_2^2\,.
\label{itsaso3}
\end{equation}
The de Sitter horizon radius of the inflating five dimensions in this
solution is twice the radius of the remaining compactified two
dimensions.  Since the field $\phi$ in this solution has the same
value, $\phi=0$, as in the monopole core, the metric~(\ref{itsaso3})
describes the entire spacetime for $\kappa\eta=\sqrt{5}$.
For higher values of $\eta$, we expect that $\phi$ remains equal to
zero, and the metric is still given by~(\ref{itsaso3}).

\section{Conclusions}

In this paper, we continued the investigation of the gravitational
field of higher-dimensional global defects that was started in
Ref.~\cite{Olasagasti}.  We have obtained numerical solutions of Einstein's
and scalar-field equations for a defect with a $(3+1)$-dimensional
core, which has a global-monopole-like field configuration in the
extra three dimensions.  We have verified that, for symmetry-breaking
scales below the higher-dimensional Planck scale, $\eta<1/\kappa$, the
spacetime of the defect worldsheet is flat, and the geometry of extra
dimensions is that of a generalized cone, in agreement with the
asymptotic solution obtained in Ref.~\cite{Olasagasti}.  The solid angle
deficit grows with $\eta$ and consumes the entire solid angle at the
critical value $\eta_c=1/\kappa$.

Our main goal in this paper has been to investigate supercritical
defects with $\eta>\eta_c$.  First, we have found that static
supercritical solutions have a curvature singularity.  It has been
known for some time that singularities of static-defect solutions in
4D can be removed by relaxing the staticity requirement and allowing
the defect cores to inflate.  We have shown that the same is true for
our 7D global monopoles.  We found numerical solutions in which the
defect core has the geometry of de Sitter space inflating at some rate
$H$.  As $H$ is increased, for each $\eta>\eta_c$, there is a
value of $H$ for which the spacetime becomes nonsingular.  The extra
dimensions of these nonsingular solutions have a ``cigar'' geometry. 

At large distances from the core, the nonsingular solutions have a
very simple form, and we have found an exact analytic solution of the
field equations describing this asymptotic behavior.  In this
solution, the five dimensional subspace, formed by the four dimensions
tangent to the defect worldsheet and the radial dimension in the extra
space, has the geometry of a 5D de Sitter space.  Thus, inflation
occurs not only along the defect, but also in one extra dimension.
(Note, however, that the radius of the defect core is smaller than
the horizon and remains fixed during the course of inflation.)  The
remaining two extra dimensions are compactified as a sphere of a
fixed radius.  Far away from the core, the scalar field approaches a
constant value $\phi_0$, which, however, is somewhat smaller than the
flat-space expectation value $\eta$.

As the symmetry-breaking scale $\eta$ is increased, the inflation rate
$H$ grows and the asymptotic scalar-field  value $\phi_0$ decreases.
For $\eta\geq\sqrt{5}\kappa^{-1}$, the scalar field vanishes in the
entire spacetime, and the defect solution turns into a vacuum solution
of Einstein equations with a cosmological term.  We interpret this as
indicating that, for higher values of $\eta$, defect solutions with a
core of a fixed transverse size do not exist.

We focused in this paper on a global-monopole-type defect in a 7D
spacetime, but since the gravitational properties of all global
defects of codimension $n>2$ are rather universal, we expect our
conclusions to apply to all such defects, with only minor
modifications.  In particular, the analogues of the $n=3$ asymptotic
solutions~(\ref{eq=phi0})-(\ref{eq=C0}) can be easily found for $n>3$.

As already noted, the inflation rate $H$ grows as $\eta$ is increased.
This behavior is opposite to what is needed for solving the
cosmological constant problem.  This indicates that global defects
cannot be used to implement the Dvali-Gabadadze-Shifman 
idea~\cite{DGS}.  
We shall discuss the gravity of higher-dimensional gauge
defects in a forthcoming paper.

When this work was in progress, we learned that Ruth Gregory has
independently studied inflating black-brane solutions.

\acknowledgements

We are grateful to Gia Dvali and Gregory Gabadadze for useful
discussions.  The work of A.V. was supported in part by the National
Science Foundation.

\clearpage

\clearpage
\begin{figure}
\begin{center}
\epsfig{file=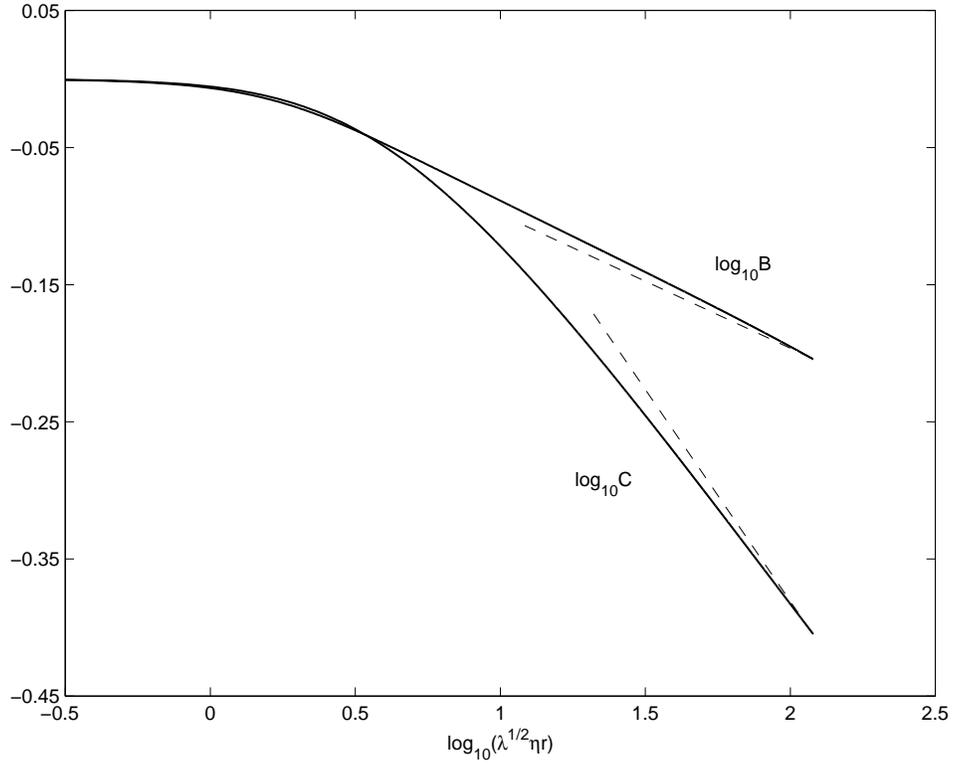,width=5in}
\end{center}
\vspace{0.5in}
\caption{
Metric coefficients $B$ and $C$ for the critical
monopole, $\kappa\eta = 1.0$.  The asymptotic power-law solutions are
shown by dashed lines.
}
\label{fig=eta1BC}
\end{figure}

\clearpage
\begin{figure}
\begin{center}
\epsfig{file=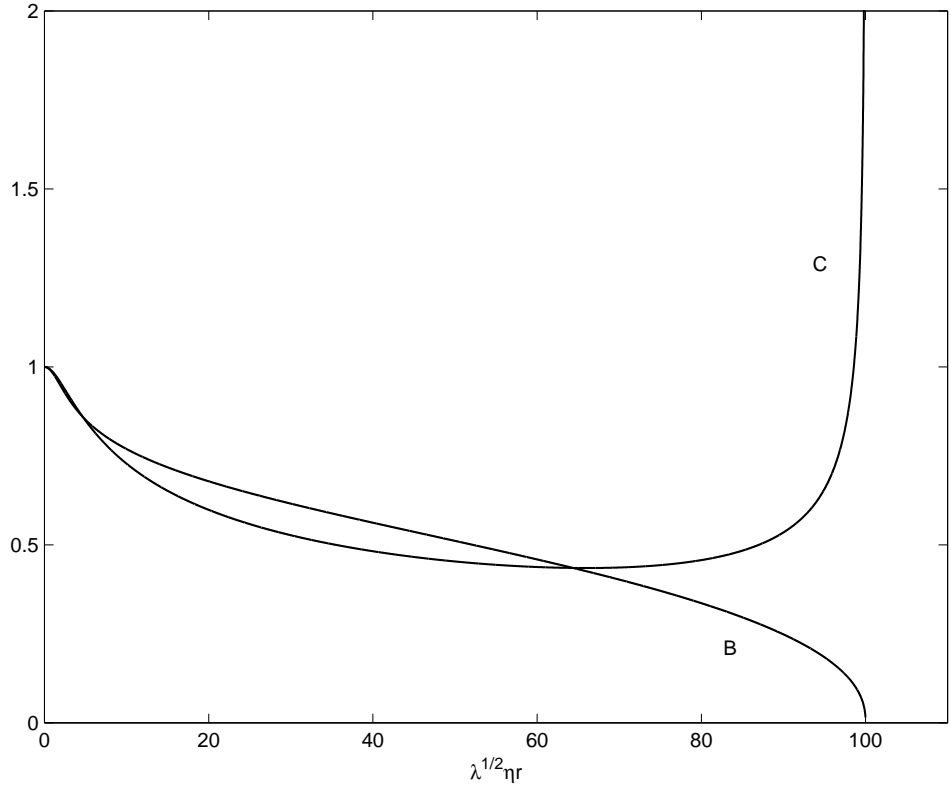,width=5in}
\end{center}
\vspace{0.5in}
\caption{
Metric coefficients $B$ and $C$ for a super-critical
monopole, $\kappa\eta = 1.06723$.  The metric becomes 
singular at a finite distance from the center.
}
\label{fig=singBC}
\end{figure}

\clearpage
\begin{figure}
\begin{center}
\epsfig{file=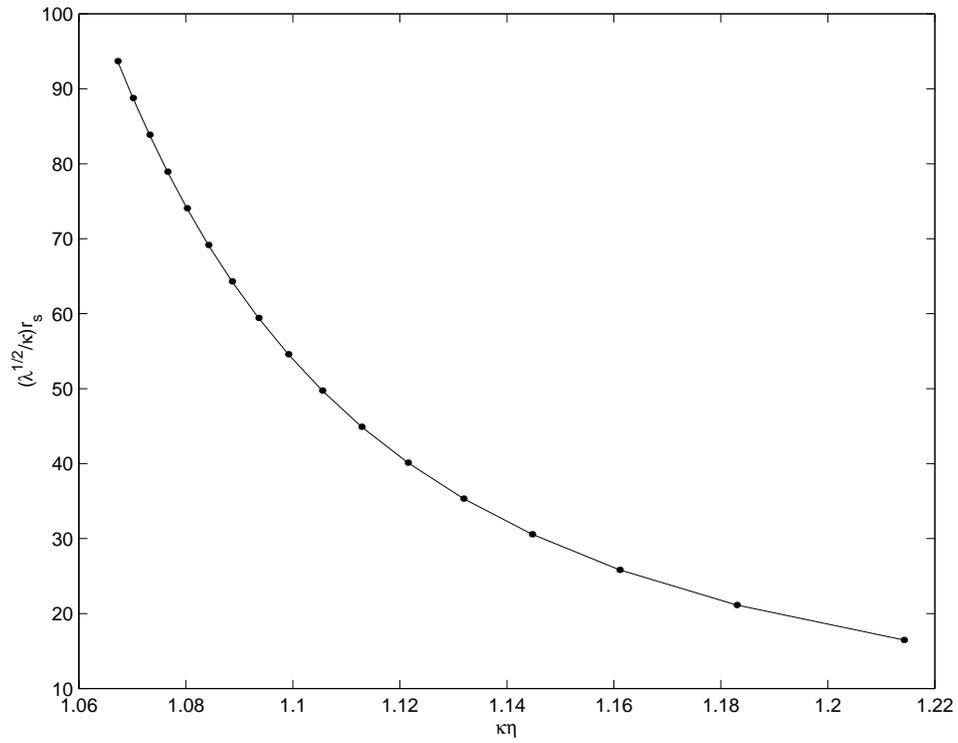,width=5in}
\end{center}
\vspace{0.5in}
\caption{
The location of the singularity $r_s$ as a function of $\eta$.
}
\label{fig=rs}
\end{figure}

\clearpage
\begin{figure}
\begin{center}
\epsfig{file=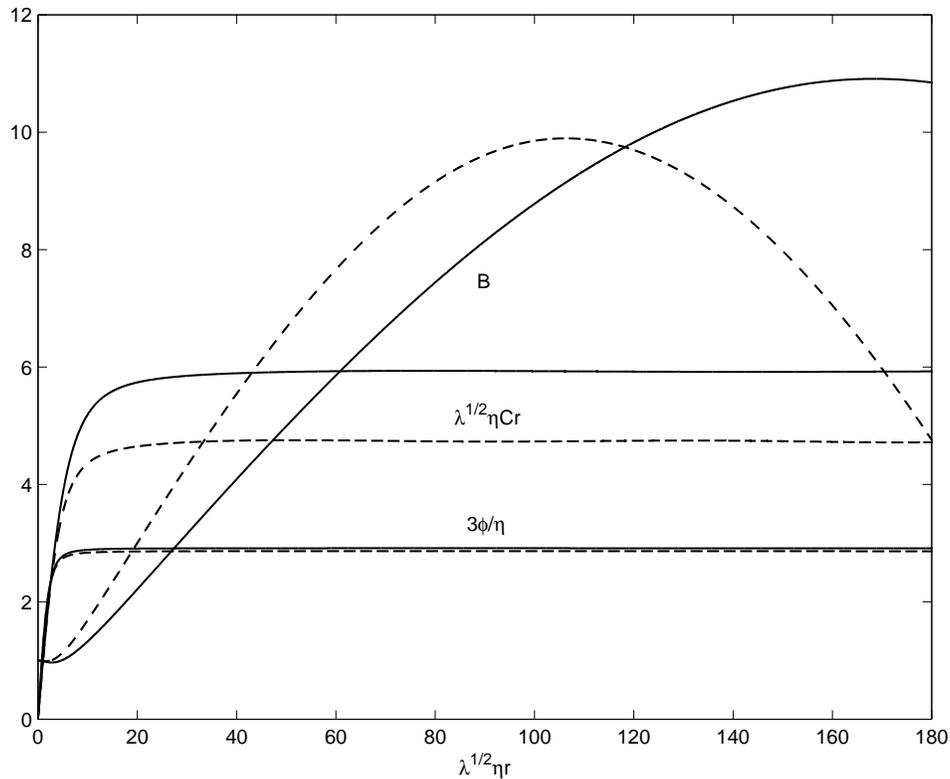,width=5in}
\end{center}
\vspace{0.5in}
\caption{
Scalar and gravitational fields of the ``cigar'' solutions with
$\kappa\eta = 1.02361$ ($H=0.10543\sqrt{\lambda}\eta$; solid line) and
$\kappa\eta = 1.03766$ ($H=0.15565\sqrt{\lambda}\eta$; dashed line).
In the asymptotic region, the scalar field and $Cr$ are nearly
constant, and $B$ behaves as a sine function.  The scalar field is
scaled to have a better discrimination.
}
\label{fig=flat}
\end{figure}

\clearpage
\begin{figure}
\begin{center}
\epsfig{file=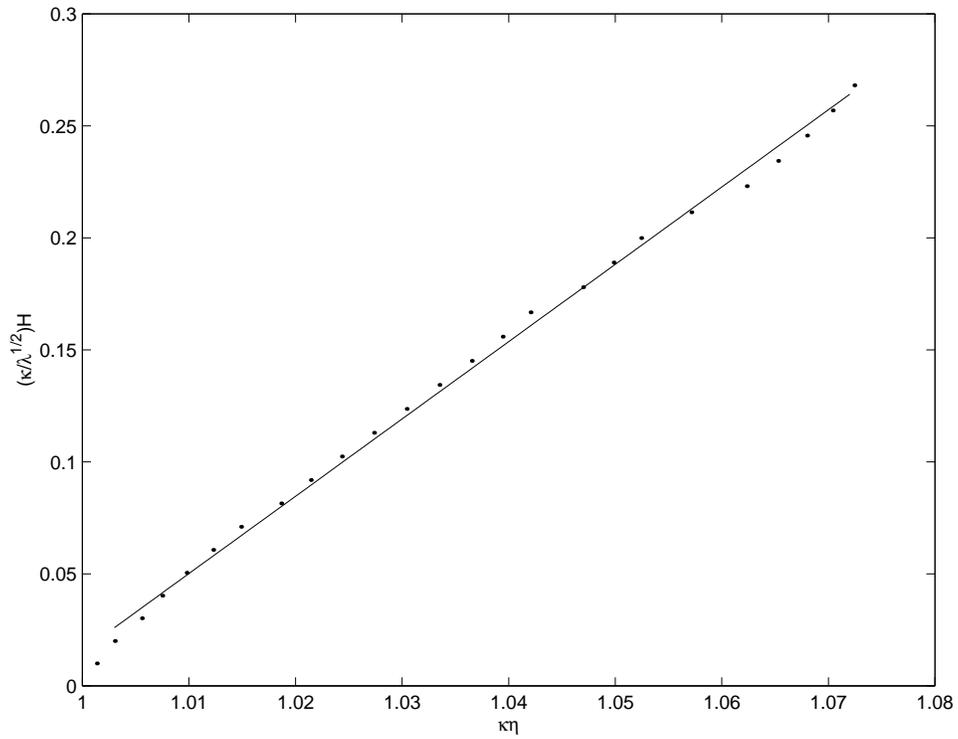,width=5in}
\end{center}
\vspace{0.5in}
\caption{
The worldsheet expansion rate $H$ as a function of $\eta$.
A linear fit is also shown.
}
\label{fig=H}
\end{figure}

\end{document}